\documentclass[aps,prb,twocolumn,showpacs,byrevtex,amsmath,superscriptaddress,floatfix]{revtex4-1}
\usepackage[dvips]{graphicx}
\usepackage{graphics}
\usepackage{epsfig}
\usepackage{amsmath}
\usepackage{amsfonts}
\usepackage{amssymb}
\usepackage{dcolumn}
\usepackage{bm}
\usepackage{longtable}
\usepackage{multirow}

\begin{document}

\title{Spectral function of few electrons in quantum wires 
and carbon nanotubes as a signature of Wigner localization}

\author{Andrea Secchi}

\affiliation{CNR-NANO S3, Via Campi 213a, 41125 Modena, Italy}

\affiliation{Dept of Physics, University of Modena and Reggio Emilia, Italy} 

\author{Massimo Rontani}
\email{massimo.rontani@nano.cnr.it.}
\affiliation{CNR-NANO S3, Via Campi 213a, 41125 Modena, Italy}

\date{\today }

\begin{abstract}
We demonstrate that the profile of the space-resolved
spectral function at finite temperature provides a signature of
Wigner localization for electrons in quantum wires and
semiconducting carbon nanotubes. Our numerical evidence is based 
on the exact diagonalization of the microscopic
Hamiltonian of few particles interacting in gate-defined quantum dots.  
The minimal temperature required
to suppress residual exchange effects in the spectral function image
of (nanotubes) quantum wires lies in the (sub-) Kelvin range. 
\end{abstract}

\pacs{73.20.Qt, 73.23.Hk, 73.21.La, 73.63.Fg}

\maketitle

After half a century of research, electrons in one dimension   
still attract attention as a paradigm of 
interacting behavior that deviates from Fermi liquid theory,
exhibiting e.g.~spin-charge separation or solid-like 
order.\cite{GiamarchiBook,Deshpande10,Ilani10}  
This impulse comes from recent experiments in  
systems with high aspect ratio---cleaved-edge overgrowth 
structures,\cite{Auslander,Steinberg06,Barak10} 
multiple-gate quantum wires,\cite{Hew08}
carbon nanotubes,\cite{Deshpande08,Kuemmeth08,Deshpande09,Churchill09,Steele09} 
nanowires\cite{Roddaro08,Kristinsdottir11}---which 
are all effectively one-dimensional as their transverse
and longitudinal degrees of freedom are decoupled.
The refinement of such devices allows to easily 
reach the dilute regime of electron density 
yet minimizing the impact of disorder.
At sufficiently low density, the Coulomb energy gain
overcomes the kinetic energy cost of localization,
hence electrons are expected to freeze their motion in space
forming a regular array---a Wigner correlated solid.\cite{Schulz93,Fiete07} 

So far, the evidence of Wigner localization
has ultimately relied on the measure of the energy 
gap between ground and low-lying excited states,
which vanishes in the dilute limit.\cite{Steinberg06,Deshpande08,Steele09,Kristinsdottir11}
This excitation energy decreases gradually
from the liquid- to the solid-like regime,
as an effect of both quantum fluctuations
and samples' finite size---systems often act as quantum
dots (QDs) in the Coulomb blockade 
regime.\cite{Steinberg06,Deshpande08,Kuemmeth08,Deshpande09,Churchill09,Steele09,Roddaro08,Kristinsdottir11}
Therefore, an alternative signature of the electron solid, 
directly related to the wave function, would be desirable. 
A possible observable is
the momentum-resolved spectral function (SF)---the quasiparticle 
wave function square modulus in reciprocal space.\cite{Rontani05,Steinberg06} 
Unpromisingly, it was predicted
that the SF was qualitatively similar in both Wigner and 
non-interacting limits\cite{Fiete05} and that   
any distinctive structure of the SF was washed out by temperature.\cite{Mueller05} 

In this Communication we demonstrate that the space-resolved spectral
function of few electrons 
provides a clear signature of Wigner localization 
at temperatures above $T_{\text{ex}}$, 
that is the characteristic scale of exchange interactions. 
This fundamental observable may be accessed through 
scanning tunneling spectroscopy (STS).\cite{WiesendangerBook,HoferRMP}
Our exact diagonalization\cite{Bryant87, Reimann02, Rontani06, Secchi09} (ED) 
results show that the SF resembles the charge density 
for $T\agt T_{\text{ex}}$, displaying 
$N$ peaks as the $N$th electron tunnels into a 
Coulomb blockaded QD already containing $N-1$ electrons.
The peak-to-valley ratio of
such image allows to assess directly the onset
of Wigner localization.
In sharp contrast, for $T\ll T_{\text{ex}}$
the SF is system-dependent and unrelated to $N$.
Overall, the joint measurements of $N$ and of the
SF are able to unveil the Wigner solid.

\begin{figure}
\makebox[\linewidth]{\unitlength=1mm
\begin{picture}(80,94)
\put(3,-3){\epsfig{file=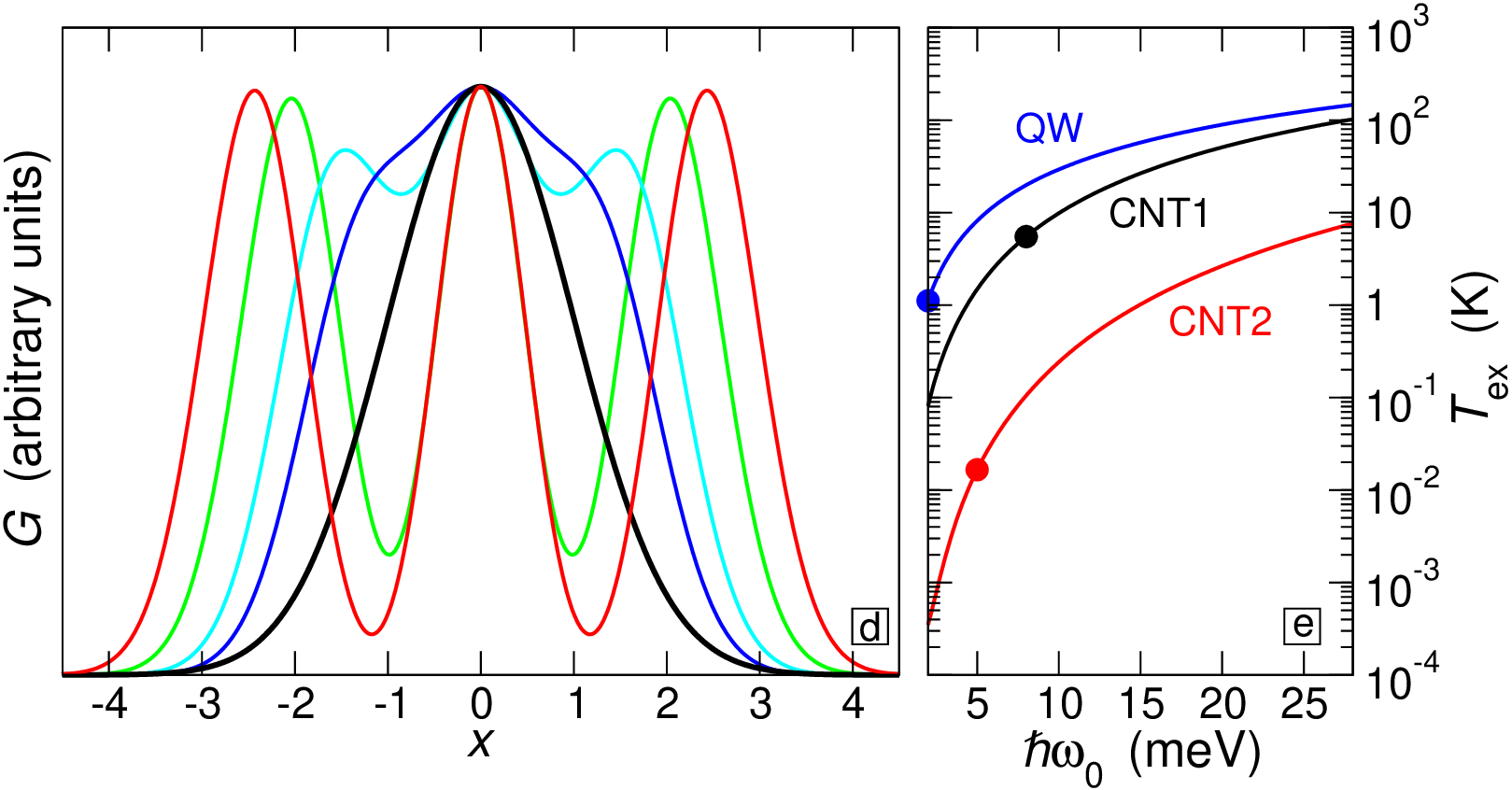,width=3.0in,,angle=0}}
\put(3,41.5){\epsfig{file=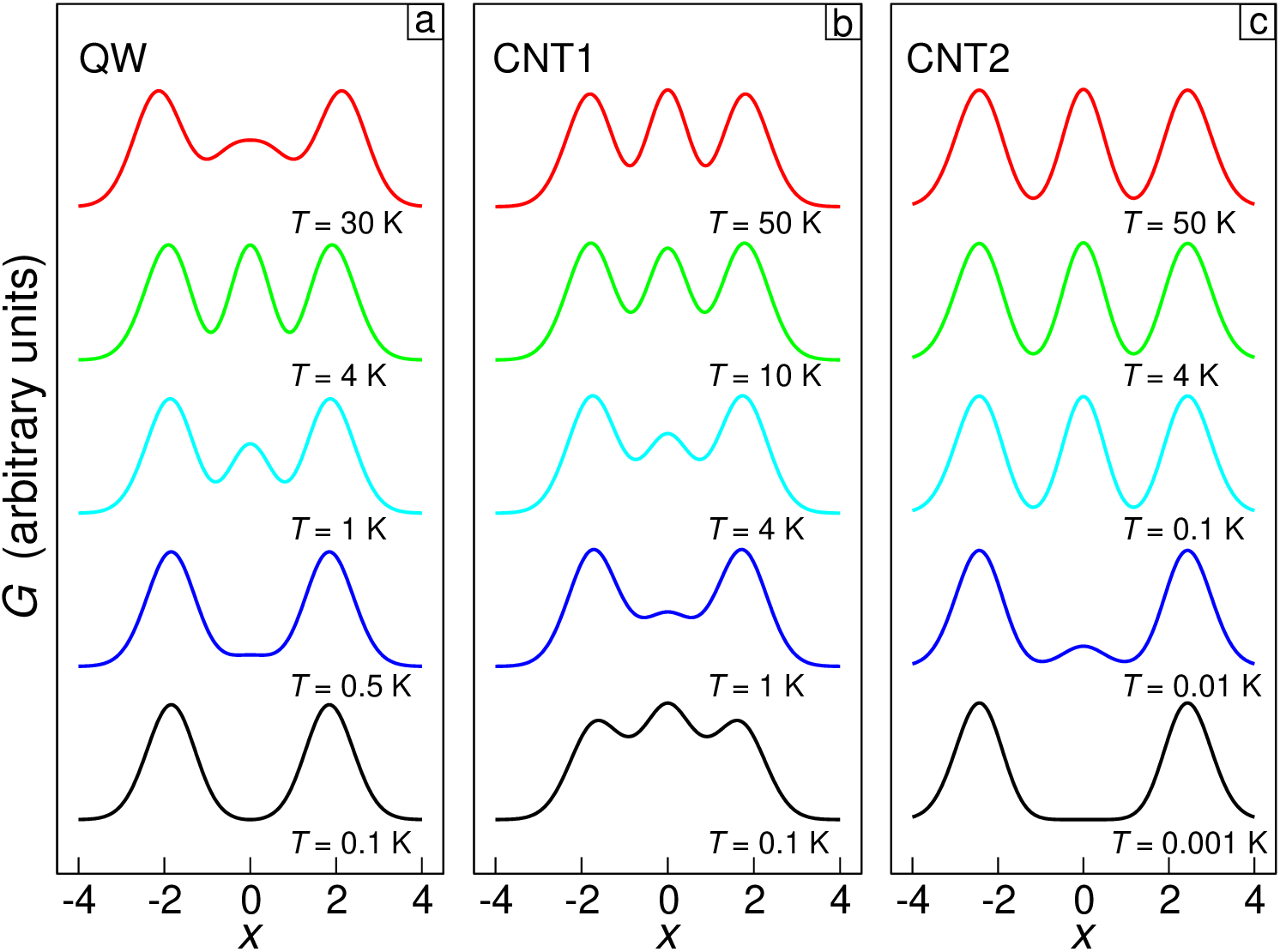,width=2.7in,,angle=0}}
\end{picture}}
\caption{(Color online) (a-c) SF signal $\cal{G}$ 
vs $x$ at various $T$ for $N=2\rightarrow N=3$.
(a) Quantum wire (QW). 
The length unit is $\ell_{\text{QD}}=$ 23.9 nm.
(b) Carbon nanotube no.~1 (CNT1). 
$\ell_{\text{QD}}=$ 26.8 nm.
(c) Carbon nanotube no.~2 (CNT2).   
$\ell_{\text{QD}}=$ 17.9 nm. 
(d) SF signal $\cal{G}$
vs $x$ at $T=$ 10 K for a CNT as $\hbar\omega_0$ is varied.
The five curves, from red [light gray] to black, correspond to 
$\hbar\omega_0=$
5, 20, 60, 100, $\infty$ meV, respectively.
The length unit is $\ell_{\text{QD}}$ and
the remaining parameters are those of CNT2. 
(e) $T_{\text{ex}}$ vs 
$\hbar\omega_0$ for QW,
CNT1, and CNT2. The filled circles 
are the values pertinent to panels (a-c).
\label{fig:all}} 
\end{figure}

The scenario reported in this Communication agrees with the theory of
the `spin-incoherent' Luttinger liquid.\cite{Fiete07}
On the other hand, crucial approximations of this theory
poorly reproduce key experimental features of
systems with a moderate number of 
electrons,\cite{Steinberg06,Deshpande08,Kuemmeth08,Deshpande09,Churchill09,Steele09,Roddaro08,Kristinsdottir11} 
and most noticeably:
(i) finite-size effects are prominent;\cite{Anfuso03,Cavaliere04,Gindikin07,Pugnetti09,Schenke09}
(ii) the occurrence of the band gap\cite{Levitov03} and
band curvature\cite{Imambekov09,Barak10}
may not be neglected.

In our ED approach we take into account all many-body correlations  
as well as the effects of 
spin-orbit coupling, valley degeneracy,
band curvature through the effective mass $m$.\cite{Secchi09}
We assume the QD confinement potential along $x$ to be harmonic, 
$V_{\text{QD}}(x)=m\omega_0^2x^2/2$,\cite{Reimann02,Secchi09} 
since this is the generic low-energy form
for gated QDs embedded in quantum wires (QWs) and semiconducting
carbon nanotubes (CNTs), as in 
Refs.~\onlinecite{Steinberg06,Deshpande08,Kuemmeth08,Deshpande09,Churchill09,Steele09}. 
We consider $N$ electrons in a QW interacting through
a screened Coulomb interaction 
$V(x,x')=e^2\epsilon^{-1}[(x-x')^2+\lambda^2]^{-1/2}$, 
with $\lambda$ being a short-range
cutoff and $\epsilon$ the dielectric constant. The CNT Hamiltonian 
is more complex, due to the presence of valleys K and K$^{\prime}$,
spin-orbit coupling, inter- and intra-valley interactions. 
The interaction potential interpolates between Coulomb and 
Hubbard-like behavior: see
Ref.~\onlinecite{Secchi09} for details.  
In all cases we diagonalize the Hamiltonian in the
space spanned by the Slater determinants built by filling with $N$
electrons the lowest 60 spin-orbitals in
all possible ways. 

The outcome of the ED consists in the $i$th excited
$N$-body states, $\left|N,i\right>$, and
their energies $E^N_i$. The SF for a given initial
state at $T=0$,
$A_{N-1,i}(x,\omega)$, is  
\begin{displaymath}
A_{N-1,i}(x,\omega)=\sum_{j}\left|\left<N,j\right|\hat{\Psi}^{\dagger}(x)
\left|N\!-\!1,i\right>\right|^2\!\!
\delta(\hbar\omega-E^N_j+E^{N-1}_i),
\end{displaymath}
with $\hat{\Psi}^{\dagger}(x)$ being the operator creating an electron 
at $x$ and $\hbar\omega$ its resonant tunneling energy, whereas  
the ground-state charge density is
$\varrho(x)=N^{-1}\left<N,0\right|\hat{\Psi^{\dagger}}(x)
\hat{\Psi}(x)\left|N,0\right>$.
Ideally, the STS differential
conductance at vanishing temperature and bias, $dI/dV$, is proportional
to $A_{N-1,0}(x,\varepsilon_F/\hbar)$, 
with $\varepsilon_F$ being the Fermi energy of the STS tip.\cite{noteSTS}
If the thermal broadening $k_BT$ is larger than typical
energy spacings,\cite{WiesendangerBook} $dI/dV$ is proportional to
\begin{equation}
{\cal{G}} =
\int d\omega \left[-\frac{\partial f(\hbar\omega)}{\partial \omega}
\right]A_T(x,\omega),
\label{eq:I}
\end{equation}
where $f(\cdot)$ is the Fermi distribution function
and $A_T(x,\omega)$ is the finite-temperature SF,
\begin{equation}
A_T(x,\omega)=\frac{1}{Z}\sum_ie^{-\beta E^{N-1}_i}A_{N-1,i}(x,\omega),
\end{equation}
with $\beta=1/(k_BT)$ and $Z=\sum_i\exp{(-\beta E^{N-1}_i)}$.
In the following we tune $\varepsilon_F$ 
appearing in $f(\cdot)$
to match the $N-1\rightarrow N$ transition between ground states.
 
\begin{figure}
\includegraphics[width=2.6in]{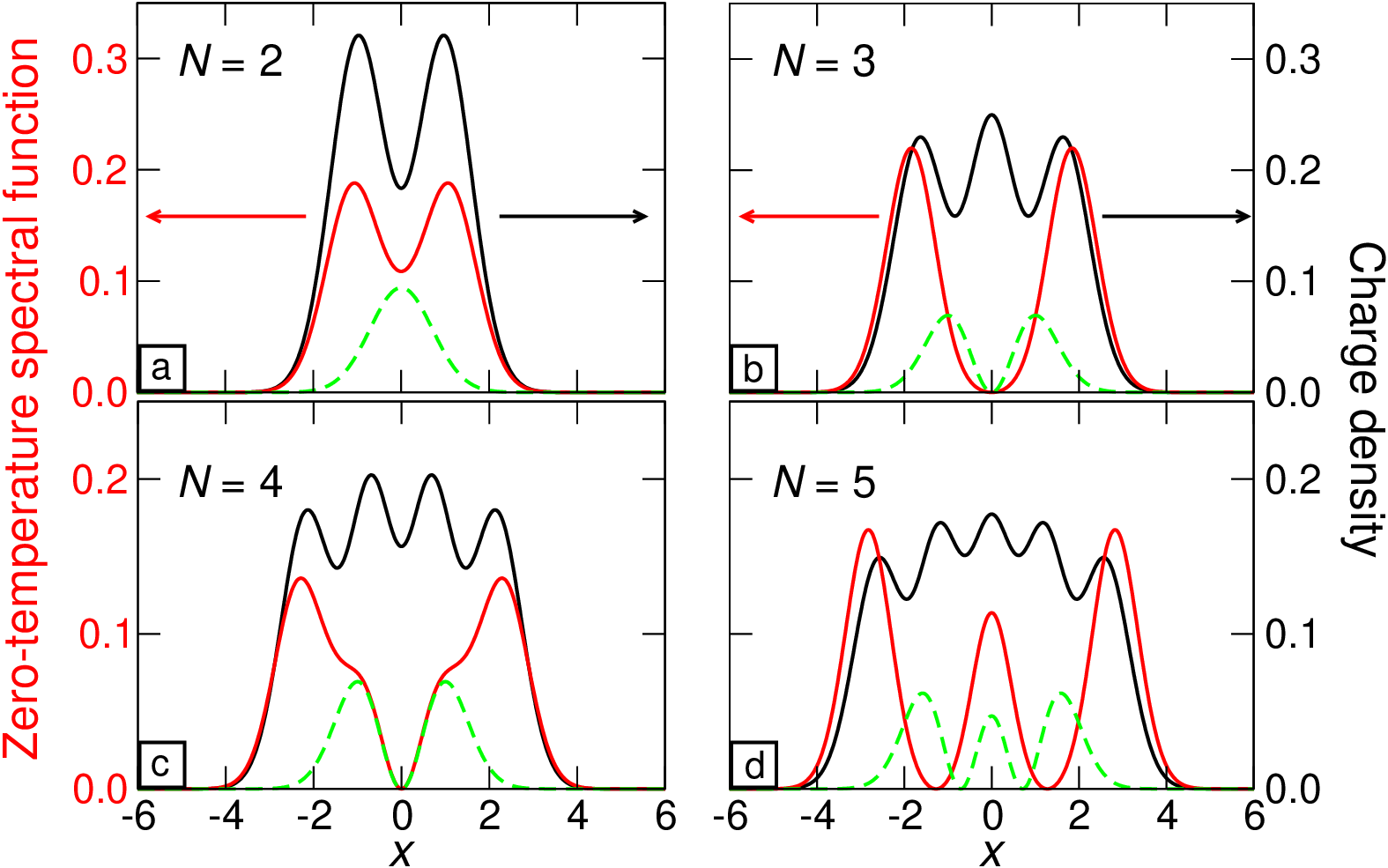}
\caption{(Color online) QW $\varrho(x)$
of $N$ electrons (black curve) and SF at $T=0$
for the $N-1\rightarrow N$ ground-state transition (solid red [gray] curve)
vs $x$ for $N=2$ (a), 3 (b), 4 (c), 5 (d).
The dashed curves are the non-interacting SFs,
rescaled by a factor 1/6. $A_{N-1,0}(x,\omega)$ is 
integrated over a tiny energy range around  
$E^N_0-E^{N-1}_0$. 
The length unit is $\ell_{\text{QD}}=$ 23.9 nm.
\label{fig:QW}}
\end{figure}

\begin{figure}
\includegraphics[width=2.7in]{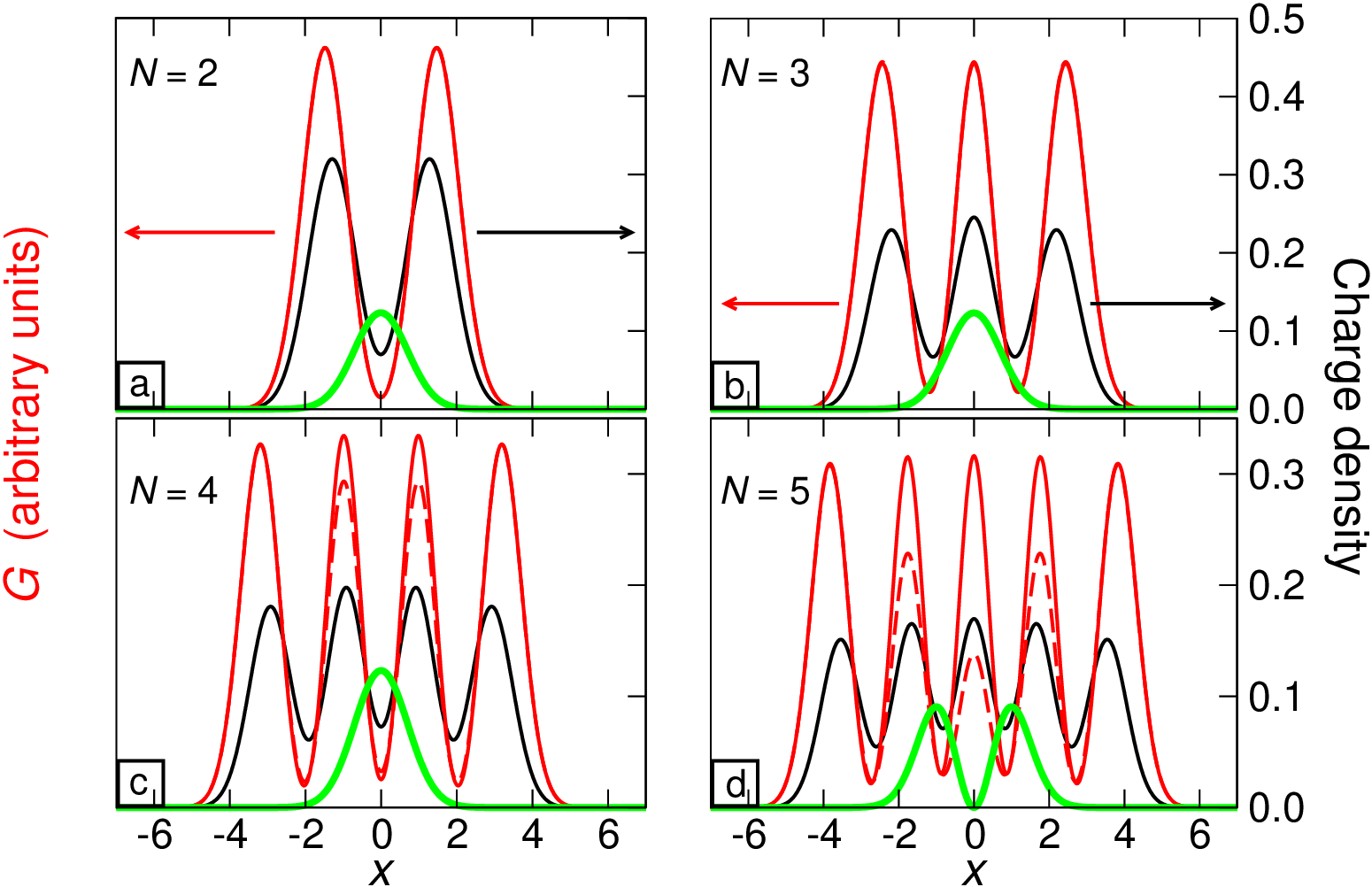}
\caption{(Color online) SF signal $\cal{G}$ 
vs $x$ at $T=0.1$ K
(dashed curve) and $T=0.5$ K (solid red [gray] and green
[light gray]
curve) for the CNT QD no. 2. The green [light gray] curve is 
the non-interacting case.  
The following $N-1\rightarrow N$ transitions
are considered: (a) $N = 2$, (b) $N = 3$, (c) $N = 4$, (d) $N = 5$.
$\varrho(x)$ for the $N$-body ground state (black curve) 
is plotted for comparison. The length unit is $\ell_{\text{QD}}=$ 17.9 nm. 
\label{fig:STS}} 
\end{figure}

To immediately grasp the key results of this Communication,
consider in Fig.~\ref{fig:all}(c)   
the SF signal $\cal{G}$ induced by the tunneling of the third electron
into a realistic CNT (here labeled CNT2)
at various temperatures.
At low temperature, $T\ll T_{\text{ex}}=$ 0.017 K, 
the signal shows two peaks
that originate from the symmetries of the ground states.
Above $T_{\text{ex}}$ the signal shows three well resolved peaks
that resemble the charge distribution of three Wigner-localized electrons
[cf.~$\varrho(x)$ in Fig.~\ref{fig:STS}(b)]. 
The peak-to-valley ratio at $T\gg T_{\text{ex}}$ directly
measures the degree of spatial localization. This is apparent
in Fig.~\ref{fig:all}(d), as the impact of few-body correlations
is reduced by increasing the 
QD confinement energy $\hbar\omega_0$ and 
hence the ratio of kinetic to Coulomb energy.  
Whereas for strong correlations (red [light gray] curve)
the three peaks are well resolved since electrons separately
localize in space, as the interaction is turned off 
$\cal{G}$ becomes featureless (black curve for $\hbar\omega_0=\infty$),
clearly discriminating
between Wigner and weakly-interacting regimes. 

Also the magnitude of $T_{\text{ex}}$ points to electron
correlation, the lower the temperature
$T_{\text{ex}}$, the stronger the localization. 
$T_{\text{ex}}$ varies significantly for typical QWs and
CNTs as a function of device parameters, like  
$\hbar\omega_0$. In Fig.~\ref{fig:all}(d)  
the increase of
$\hbar\omega_0$ quenches correlations and
amplifies the effects of Fermi statistics,
raising $T_{\text{ex}}$. 
For realistic parameters [cf.~circles in Fig.~\ref{fig:all}(d)],
we find that $T_{\text{ex}}$ may be as low as 10 mK 
in some semiconducting CNTs.
On the other hand, $T_{\text{ex}}$ is one-two orders of magnitude
higher in quantum wires (QWs), $T_{\text{ex}} \sim 1$ K, 
as an effect of the different impact of screening.

Figure \ref{fig:QW} shows $\varrho(x)$ and the zero-temperature SF 
up to five electrons for a typical wire QD. 
In the ED we chose $\lambda=$ 5 nm, bulk GaAs parameters, 
and $\hbar\omega_0=$ 2 meV as single-particle energy spacing,
providing a characteristic QD length
$\ell_{\text{QD}}=(\hbar/m\omega_0)^{1/2}=23.9$ nm. As it is seen from the
spread of $\varrho(x)$ along the axis
(black curves in Fig.~\ref{fig:QW}), this corresponds to a typical 
size of $\approx$ 200 nm for $N=5$, which is comparable to the size
$L_{loc}=$ 230 nm of the QD in the dilute limit of 
Ref.~\onlinecite{Steinberg06} (Table I).
The profiles of $\varrho(x)$ 
point to the partial localization of the $N$ electrons as $N$ peaks
emerge from a featureless liquid droplet. On the other hand, the 
SFs for the $N-1\rightarrow N$ ground state transitions
(solid red [gray] curves 
in Fig.~\ref{fig:QW}) are qualitatively distinct from $\varrho(x)$
at given $N$. This is patent for $N=3,4,5$, 
with SFs displaying one, one, and two nodes, respectively,
whereas the corresponding $\varrho(x)$'s have no nodes. 
As seen in Fig.~\ref{fig:all}(a),
${\cal{G}}$ is insensitive to temperature in the range
$T\ll T_{\text{ex}}=$1.1 K, with  
$k_B T_{\text{ex}}$ being the energy splitting between
the lowest two-electron triplet and singlet states.
Here $k_B T_{\text{ex}}$ provides the energy scale of
low-lying excitations.

The SFs appearing in Fig.~\ref{fig:QW} are similar 
to those obtained in the absence of interaction (dashed lines). 
In the non-interacting limit the SF is the square modulus of 
the orbital occupied 
by the $N$th electron that enters the QD.\cite{Rontani05} 
Such orbital has zero, one, one, and
two nodes for $N=2,3,4,5$, respectively, since electrons 
fill in each orbital level twice due to Kramers degeneracy.  
As the symmetries of the quantum states
do not change in the considered range of interaction,
no qualitative differences are seen for the interacting SFs.\cite{Fiete05} 

\begin{figure}
\includegraphics[width=2.7in]{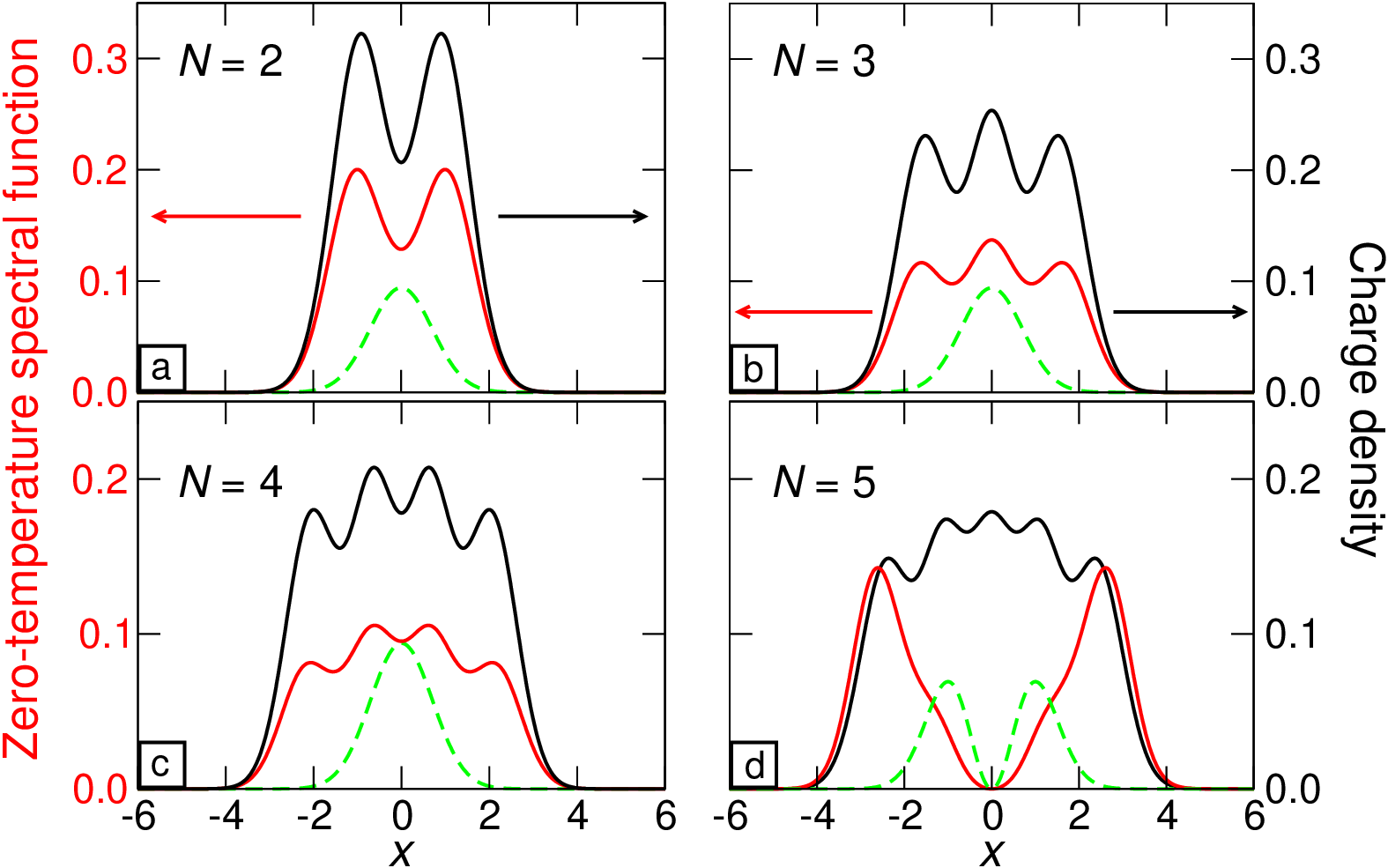}
\caption{(Color online) CNT  
$\varrho(x)$ of $N$ electrons (black curve) and SF at $T=0$ 
for the $N-1\rightarrow N$ ground-state transition (solid red [gray] curve)
vs $x$ for $N=2$ (a), 3 (b), 4 (c), 5 (d).
The dashed curves are the non-interacting SFs,
rescaled by a factor 1/6.
$A_{N-1,0}(x,\omega)$ is 
integrated over a tiny energy range around 
$E^N_0-E^{N-1}_0$.
The length unit is $\ell_{\text{QD}}=$ 26.8 nm.
\label{fig:CNT}}
\end{figure}

In the following we discuss two examplar CNT cases.
Figure \ref{fig:CNT} is the analogue of Fig.~\ref{fig:QW} for the 
CNT device investigated in Ref.~\onlinecite{Kuemmeth08}
(here labeled as no. 1).
The ED parameters ($\hbar\omega_0=8$ meV, $\epsilon=3.5$, radius $R=3.6$ nm) 
were chosen in order
to reproduce the measured chemical potentials,  
as detailed in Ref.~\onlinecite{Secchi09}.
Apart from length renormalization ($\ell_{\text{QD}}=$ 26.8 nm),
charge densities (black curves in Fig.~\ref{fig:CNT}) are 
similar to those of the QW (Fig.~\ref{fig:QW}).
On the contrary, the zero-temperature SFs for the CNT 
(solid red [gray] curves 
in Fig.~\ref{fig:CNT}) are drastically different from those
of Fig.~\ref{fig:QW}, being all nodeless except for the 
$N=4\rightarrow N = 5$ transition. 
This trend is qualitatively similar
to the non-interacting filling sequence (dashed lines 
in Fig.~\ref{fig:CNT}), as each CNT level
is four-fold degenerate in the absence of spin-orbit coupling,
due to both spin and valley degeneracies. The spin-orbit interaction
splits the multiplet into two doublets (here separated
by $\Delta E_{\text{SO}} =$ 0.367 meV) but leaves the
spin-orbitals unchanged. Since 
$\Delta E_{\text{SO}}$ remains 
the energy scale of the low-lying excitations even in the presence
of interactions in the sample no.~1 (Refs.~\onlinecite{Kuemmeth08, Secchi09}), the SFs of 
Fig.~\ref{fig:CNT} are unaffected by temperature for 
$T\ll \Delta E_{\text{SO}}/k_B\sim 4$ K 
[cf.~Fig.~\ref{fig:all}(b)]. 

The SFs of Figs.~\ref{fig:CNT}(a), (b), and (c) 
display $N$ peaks as the $\varrho(x)$'s, and
placed approximately in the same locations.
This genuine effect of interaction, reminescent
of the partial Wigner localization occuring in the QD,
takes place also in the QW [Fig.~\ref{fig:QW}(a)] 
and it has been observed for
the tunneling of the second electron into
elongated self-assembled InAs QDs.\cite{Maruccio07}

The second CNT QD that we study experiences stronger
interactions than the first one, as
an effect of the smaller energy spacing ($\hbar\omega_0=$ 5 meV),
dielectric screening ($\epsilon=2.5$), radius ($R=1$ nm). 
In terms of parameters, the CNT QD no.~2 lies 
in the middle between the devices investigated in 
Refs.~\onlinecite{Deshpande08} and \onlinecite{Kuemmeth08}
(see Ref.~\onlinecite{Secchi09} for their placement 
in a phase diagram). As it is shown in Fig.~\ref{fig:STS}
(black curves), the peak-to-valley ratios of charge densities
are about twice as large as those in Figs.~\ref{fig:QW} and
\ref{fig:CNT}, hence electrons are strongly localized.  

\begin{figure}
\includegraphics[width=2.5in]{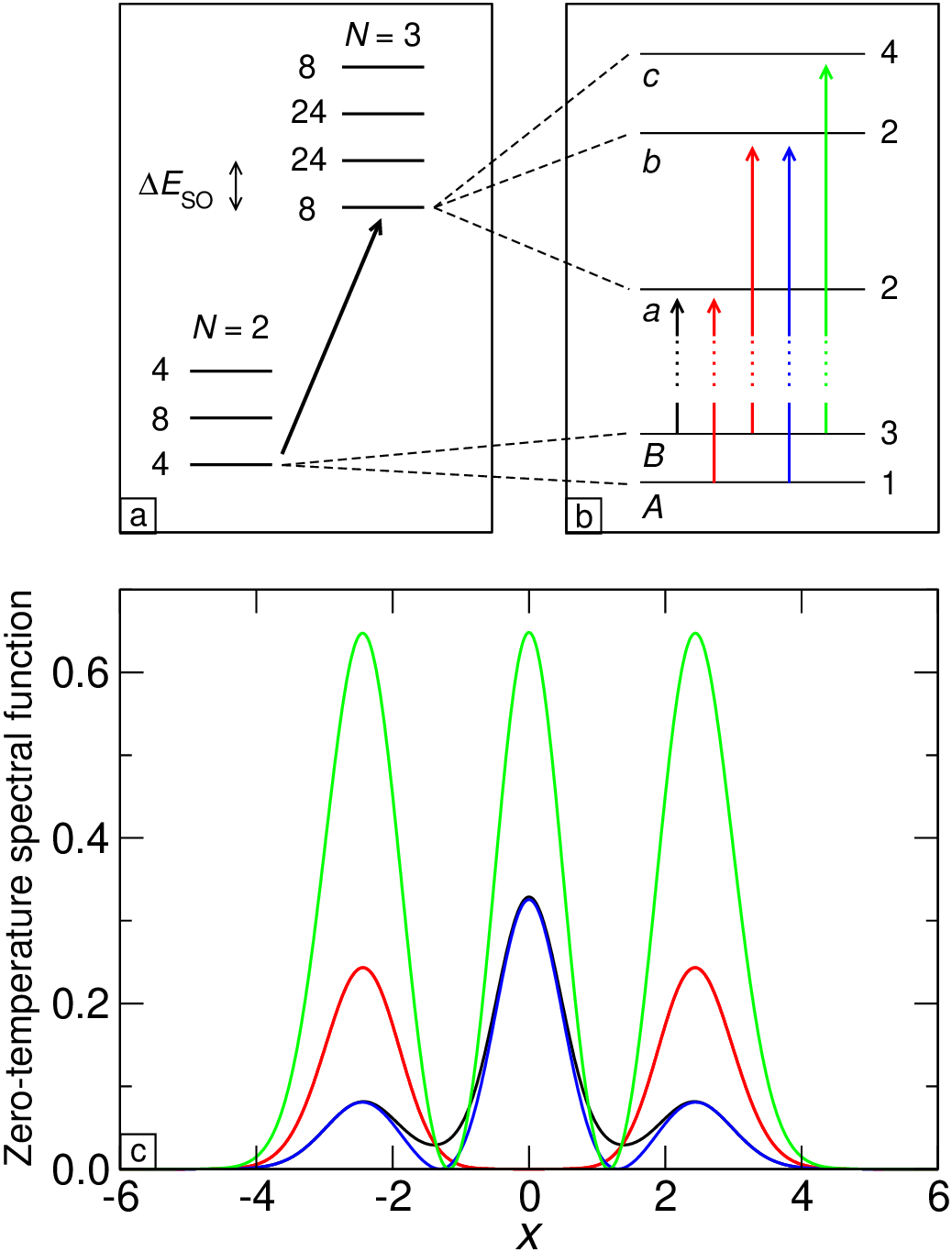}
\caption{(Color online) SFs of the CNT QD
no.~2. (a) Lowest excited states for $N=2$
and $N=3$. The numbers label sub-multiplet
degeneracies. (b) Blow-up of panel a (not in scale).  
Arrows point to 
allowed tunneling transitions. (c) SFs of allowed
transitions vs $x$ at $T=0$. 
The various $A_{N-1,0}(x,\omega)$ are 
integrated over tiny energy ranges around the resonance energies
highlighted by the arrows in panel b, 
according to the color [gray tone] code.   
The length unit is $\ell_{\text{QD}}=$ 17.9 nm.
\label{fig:scheme}}
\end{figure}

In this Wigner regime the low-lying excited
states are easily thermally populated and are highly degenerate. 
The reason is that exchange interactions between localized
electrons are suppressed ($T_{\text{ex}}$ is only 17 mK), 
therefore each electron may flip its
spin independently from the others at low energy 
cost.\cite{Secchi09}
This is true also for the isospin---the orbital angular momentum
along $x$ labelling valleys K and K$^{\prime}$.
The overall result is that the many excited states which differ only
in the (iso)spin value are almost degenerate.
This situation is illustrated in Fig.~\ref{fig:scheme}(a),
where we depict the excited-state ladders for $N=2$ 
and $N=3$. In the ultimate Wigner limit one
expects the ground state to be exactly $4^N$ degenerate, as each
electron may flip its spin and isospin in four different ways. 
The effect of spin-orbit interaction 
(here $\Delta E_{\text{SO}}=$ 1.32 meV)
is to split the ground-state multiplet into equally spaced 
sub-multiplets 
(three and four 
for $N=2$ and $N=3$, respectively).

Residual exchange interactions further split the levels, 
as illustrated by the blow-up of the two
lowest sub-multiplets for $N=2$ and $N=3$ shown in
Fig.~\ref{fig:scheme}(b) (not in scale).
For example,
states A and B for $N=2$, having the 
orbital part of the wave function even and odd 
with respect to reflection symmetry, respectively,  
by definition are split by $k_BT_{\text{ex}}=$ 1.38 $\mu$eV. Similarly,
states $a$, $b$, $c$ for $N=3$ have distinct orbital
symmetries and are split by 4.45 $\mu$eV ($a$ from $b$) and
2.26 $\mu$eV ($b$ from $c$), respectively. 

It is clear that already at $T_{\text{ex}} = 17$ mK the excited 
states of Fig.~\ref{fig:scheme}(b) are significantly populated.
Therefore, even at low temperatures---say
$T=$ 100 mK for state-of-the-art STS---the SF is a statistical 
mixture of several transitions, explicited in Fig.~\ref{fig:scheme}.
The arrows depicted in Fig.~\ref{fig:scheme}(b) point to
the low-energy transitions between $N=2$ and $N=3$
that are allowed by spin, isospin, and orbital symmetries.\cite{Secchi09}
The corresponding SFs at $T=0$ are shown in
Fig.~\ref{fig:scheme}(c). Each plot
of Fig.~\ref{fig:scheme}(c) maps onto a different transition
[the arrow of like color (gray tone) in Fig.~\ref{fig:scheme}(b)] 
except the red [light gray] plot which is identical for 
both $A\rightarrow a$ and $B\rightarrow b$ resonances.  
 
The SFs plotted in Fig.~\ref{fig:scheme}(c) 
differ among themselves with regards to both the
number of peaks and their relative intensities.
The variations are dictated by the symmetries
of initial and final states involved in the tunneling
transition.
For example, the fundamental transition $A \rightarrow a$
between two- and three-electron ground states displays two peaks 
located at opposite positions (red [light gray] curve), 
whereas all other depicted SFs have three peaks each. 
This shows that the 
SF at $T=0$ may greatly deviate from the charge-density profile,
corroborating the findings of Figs.~\ref{fig:QW} and \ref{fig:CNT}.
On the other hand, the locations of the maxima of 
curves in Fig.~\ref{fig:scheme}(c) coincide. 

By statistically averaging the SFs of low-lying transitions,
as those shown in Fig.~\ref{fig:scheme}(c), one obtains the 
finite-temperature signal $\cal{G}$ [cf.~Eq.~(\ref{eq:I})]. 
Figure \ref{fig:STS} shows the pattern of
$\cal{G}$ at $T=$ 0.1 K (dashed curves)
and $T=$ 0.5 K (solid red [gray] curves)
for transitions $N-1\rightarrow N$
up to $N=5$. The small dependence of $\cal{G}$ on the
temperature exhibited in Figs.~\ref{fig:STS}(c-d)
[see differences between
dashed and solid red (gray) curves] is a finite-size effect
due to the form of the potential $V_{\text{QD}}(x)$, as the
electron density slightly increases with $N$ (Ref.~\onlinecite{Reimann02}).
Remarkably, $\cal{G}$ has a regular behavior as a function of $N$
already at $T=0.5$ K,  
systematically displaying $N$ peaks of comparable heights 
whose positions are close to (but non coinciding with) the locations
of the maxima of $\varrho(x)$ (black curves).  

The pattern of $\cal{G}$ shown in Fig.~\ref{fig:STS}---exhibiting
high peak-to-valley ratio---is 
peculiar of the Wigner regime and should be contrasted with the
featureless, non-interacting profile (green [light gray] curves). 
Indeed, the low-lying excited states,
as those of Fig.~\ref{fig:scheme}(a), have all roughly
the same orbital wave function modulus, similar to the vibrational 
wave function of nuclei of polyatomic molecules.\cite{Secchi09}  
The differences among orbital states, as well as those
among SFs [cf.~Fig.~\ref{fig:scheme}(c)], originate
from the different nodal surfaces.
Since the weight is mainly localized around 
the equilibrium positions of electrons, nodeless interstitial regions
may hardly be distinguished from nodal regions. 
Therefore, the statistical average of excited states shows 
a regular trend, linked to the positions of localized electrons.

At sufficiently high temperatures, the SF signals $\cal{G}$
of all investigated devices behave similarly.
This is shown in Fig.~\ref{fig:all} 
for the tunneling transition $N=2\rightarrow N=3$.
Above their respective temperatures $T_{\text{ex}}$,
highlighted as circles in Fig.~\ref{fig:all}(d),
all $\cal{G}$ profiles exhibit three
peaks of similar height: see
the curves for $T=4$, 10, 0.1 K
in Figs.~\ref{fig:all}(a), (b), (c), respectively.
The different peak-to-valley ratios of these curves
measure the degree of Wigner
localization, the lower $T_{\text{ex}}$,  
the higher the ratio.

At $T \gg T_{\text{ex}}$ 
the central peak of the QW signal 
is depleted again [red (light gray) 
curve for $T=30$ K in Fig.~\ref{fig:all}(a)],
whereas CNT profiles 
[red (light gray) curves in Figs.~\ref{fig:all}(b) and (c)]
remain stable well above 50 K.
This change is due to the excitation of the energy scale
associated to charge. In fact,
the charging energy of the QW, estimated as the energy difference
between the resonance energies of the first two electrons, is 
$5.1$ meV,  that is comparable with 30 K, whereas CNT charging 
energies are much larger ($\sim 20$ meV).  

In conclusion, we have shown that the spatial dependence
of the spectral function provides a clear fingerprint for Wigner 
localization, as the temperature overcomes the energy scale of 
exchange interactions. This temperature is low enough in 
both semiconducting carbon nanotubes and quantum wires to make 
scanning tunneling spectroscopy feasible. 
This effect has not been seen in past experiments,\cite{Venema99,Lee04} 
likely due to metallic screening, as well as
to the presence of disorder and
scattering from boundaries. 
We hope our prediction may stimulate further 
work along this path. 

\begin{acknowledgments}
We thank Shahal Ilani, Vikram Deshpande, Guido Goldoni,
and Giuseppe Maruccio for stimulating discussions.
This work is supported by projects MIUR-FIRB no. RBIN04EY74,
Fondazione Cassa di Risparmio di Modena `COLDandFEW',
CINECA-ISCRA nos. HP10BIFGH8 and HP10C1E8PI. 
\end{acknowledgments}


\begin{thebibliography}{35}
\bibitem{noteSTS}
The STS tip dynamically alters the local potential seen   
by electrons in the dot,\cite{Fogler06,Qian10,Westervelt11}
possibly affecting Wigner-like correlations,  
hence it is crucial to make the measurement weakly invasive.   
This may be achieved by maximizing the tip-dot
distance and minimizing the bias voltage, which sets
the extrinsic limit of the measurement to
the sensitivity to $dI/dV$. The intrinsic limit is the condition that  
the tip spatial resolution, being proportional to the tip-dot distance,
is smaller than the average
separation between localized electrons.\cite{Fogler06}
\end{thebibliography}
\end{document}